\def \aap {\textit{A\&A} }

\documentclass{iaus}
\usepackage{graphicx}

\title[Identifying Transiting Circumbinary Planets] 
{Identifying Transiting Circumbinary Planets}

\author[Aviv Ofir]   
{Aviv Ofir$^1$}

\affiliation{$^1$School of Physics and Astronomy, Raymond and Beverly Sackler Faculty of Exact Sciences, Tel Aviv University, Tel Aviv, Israel \\ email: {\tt avivofir@wise.tau.ac.il}}

\pubyear{2008}
\volume{253}  
\pagerange{1--4}
\jname{Transiting Planets}
\editors{A.C. Editor, B.D. Editor \& C.E. Editor, eds.}
\begin{document}

\maketitle

\begin{abstract}
Transiting planets manifest themselves by a periodic dimming of their host star by a fixed amount. On the other hand, light curves of transiting circumbinary (CB) planets are expected to be neither periodic nor to have a single depth while in transit, making BLS [Kov{\'a}cs et al. 2002] almost ineffective. Therefore, a modified version for the identification of CB planets was developed - CB-BLS. We show that using CB-BLS it is possible to find CB planets in the residuals of light curves of eclipsing binaries (EBs) that have noise levels of 1\% or more. Using CB-BLS will allow to easily harness the massive ground- and space- based photometric surveys to look for these objects. Detecting transiting CB planets is expected to have a wide range of implications, for e.g.: The frequency of CB planets depend on the planetary formation mechanism  - and planets in close pairs of stars provides a most restrictive constraint on planet formation models. Furthermore, understanding very high precision light curves is limited by stellar parameters - and since for EBs the stellar parameters are much better determined, the resultant planetary structure models will have significantly smaller error bars, maybe even small enough to challenge theory.
\keywords{methods: data analysis – binaries: eclipsing – planetary systems – occultations -  binaries : close}
\end{abstract}

\firstsection 

\section{The Challenge}

\begin{figure}[b]
\begin{center}
 \includegraphics[width=5in]{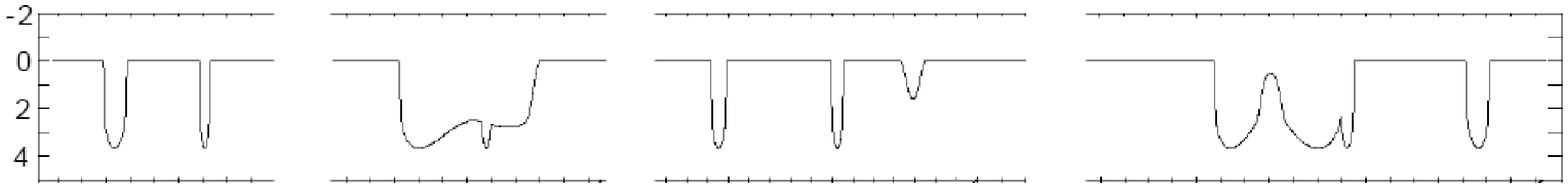} 
 \caption{Model LCs of a transiting CB planet (taken from Deeg \etal 1998). The leftmost panel shows the normal case, with two short and separated transits. The second panel shows a planetary transits near a binary mutual eclipse - producing a long transit. The two rightmost panels show transits of a planet with a long period: such planets have a transversal velocity that is slower than the binary
components, and multiple transits with complicated shapes are more likely to occur.}
\label{CB_LC}
\end{center}
\end{figure}

Light curves (LCs) of transiting planets around single stars have precise depths, durations and epochs - and changes in any of these tell us something about the system. Light curves of transiting CB planets are expected to have very different photometric and temporal characteristics (Fig. \ref{CB_LC}), where none of these properties hold. In general, photometric signal = blocked flux / total flux, but for CB planets neither is constant: the total flux changes all the time because of ellipsoidal variation and binary eclipses. The blocked flux varies according to the surface brightness of the binary component being transited. The temporal characteristics arise from the fact the transits are not periodic, and that their duration is highly variable - depending more on the transversal velocity of the binary components than on the planetary motion (the binary components can move either in parallel or anti-parallel to the planet's motion, enabling long and short transits, respectively).

\section{Solution: CB-BLS}

We assume the EB is solved as usual ($P_B, T_0, e, \omega, i_B, R_{1,2}$), and normalize the LC and EB model to max(model flux)=1. We then first deal with the photometric characteristics and then with the temporal characteristics.

\underline{Photometric characteristics:} regularizing all depths is done simply by multiplying each point on the residuals LC by it's corresponding model value - making all transits well-defined as the depth against full binary flux. We define the CB-BLS statistic by generalizing the BLS statistic to a two-box fit to allow for different surface brightness of the binary components:
\begin{equation}
\frac{s^2}{r(1-r)} \rightarrow \frac{(s_1+s_2)^2}{1-(r_1+r_2)}+\frac{s^2_1}{r_1}+\frac{s^2_2}{r_2}
\end{equation}

Where the left side is the usual BLS statistic and the right side is the new CB-BLS statistic. Similarly to BLS, $s_1$ and $r_1$ are a weighted sum and a sum of weights of points in-transit of component 1, and similarly for $s_2$, $r_2$.

\begin{figure}
\begin{minipage}[t]{0.5\linewidth} 
\centering
\includegraphics[width=6cm]{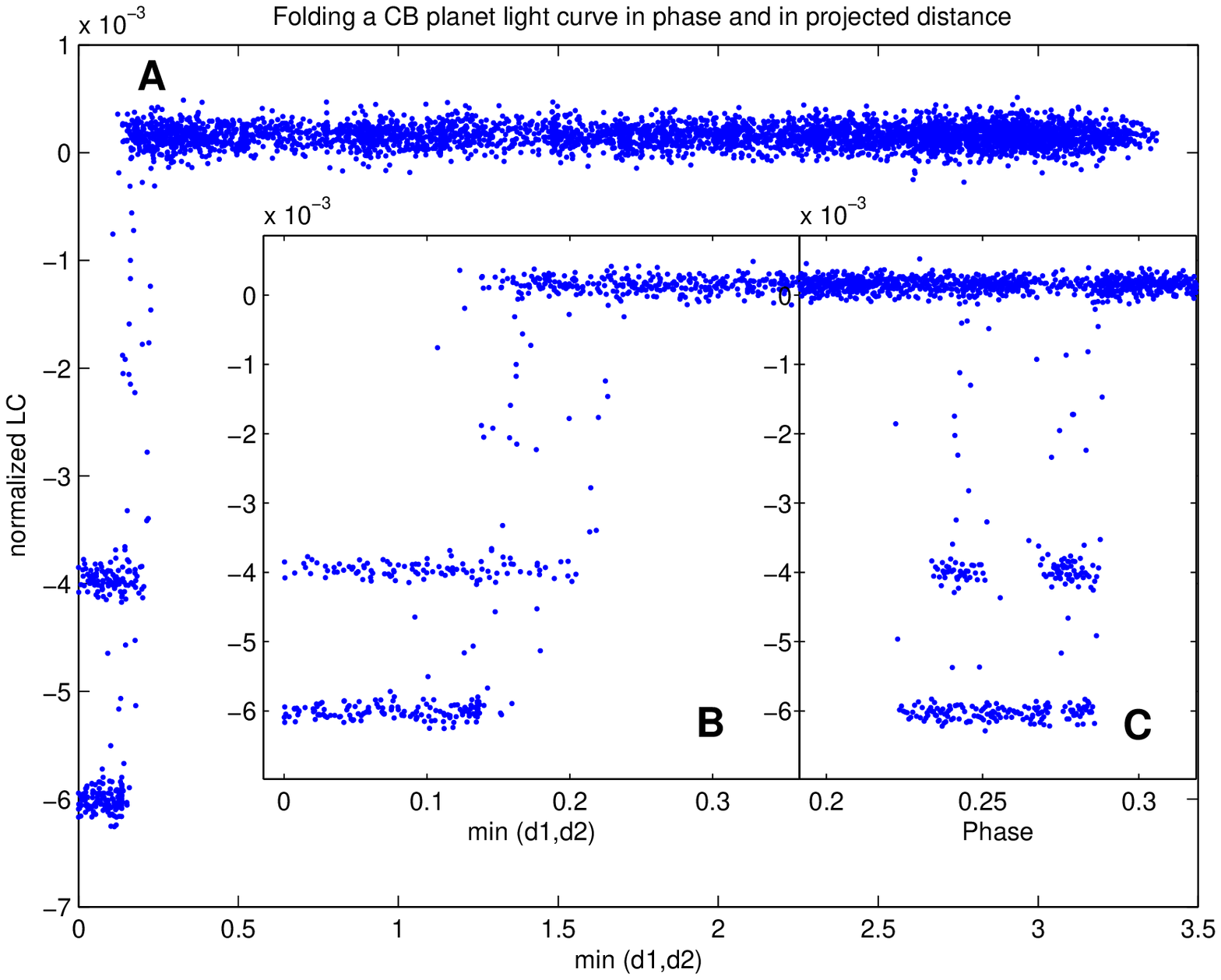} 
\caption{Fitting orbital models to the LC allows to "fold" the it in projected distance from the binary components (labeled as d1, d2). The left panel (A, and zoomed-in in B) display such a folding of a simulated LC with very low white noise of only 0.01\% to aid visibility. The LC is plotted against min(d1,d2) as derived from one model (only the half of the points where the planet is in front of the stars are shown). Evidently, In-transit points are well-separated from out-of-transit points. The different surface brightness and sizes of the stars mean different depths and distances where transits begin to occurs, respectively. For comparison, a simple phase-folding of the same data is given in the right panel (C) showing that in- and out of- transit points are not well separated, significantly reducing the detectability of the signal.}
\label{LC_vs_d}
\end{minipage}
\hspace{0.5cm} 
\begin{minipage}[t]{0.5\linewidth}
\centering
\includegraphics[width=6cm]{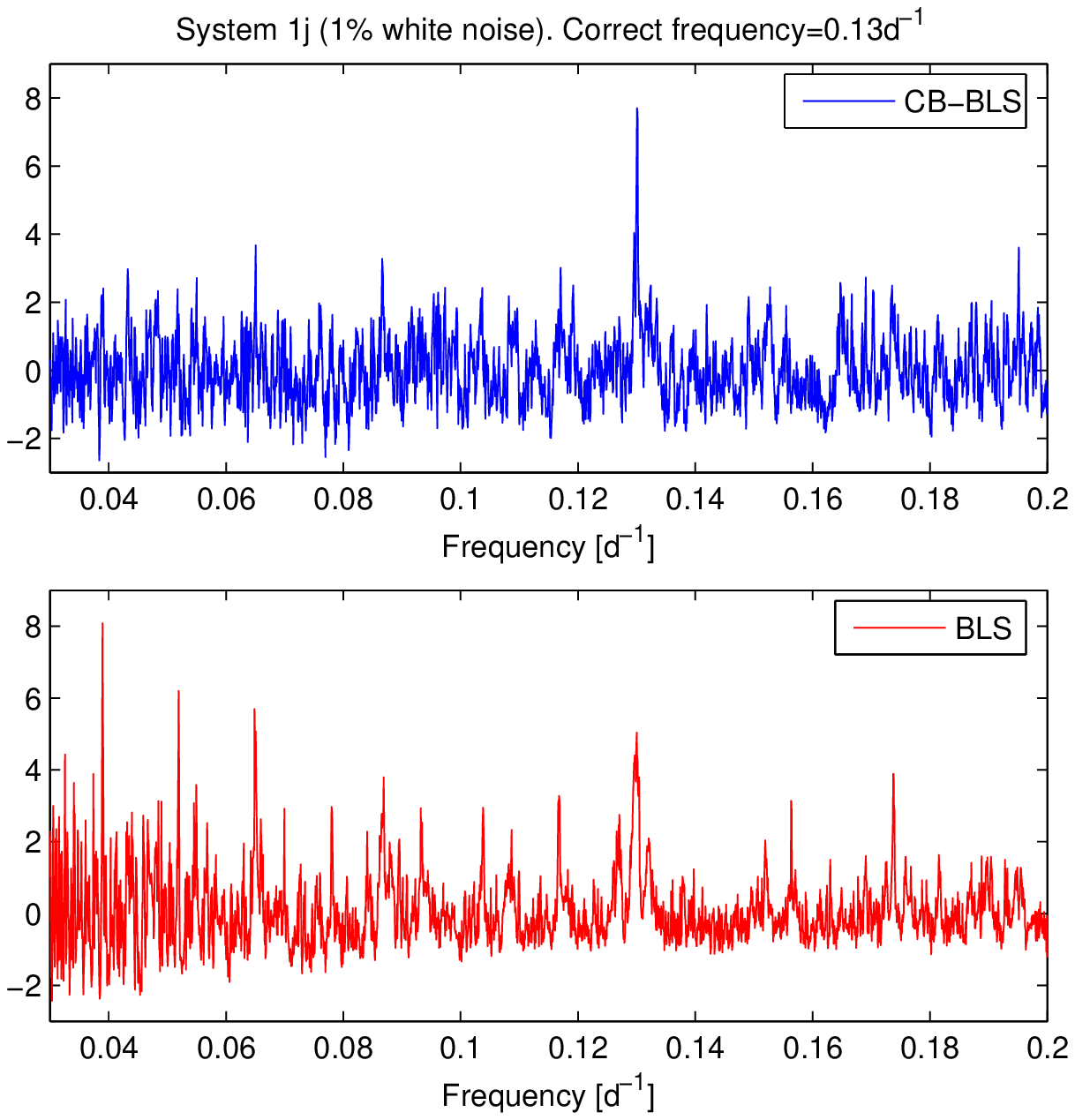} 
\caption{The fraction of correct identifications vs. the amplitude of added white noise for CB-BLS and BLS. 50 realizations were used at each noise level. Note the simulated transits were 0.4\% and 0.6\% deep}
\label{OnePeriodogram}
\end{minipage}
\end{figure}

\underline{Temporal \rightarrowfill Geometric characteristics:} transits are not a function of time, but of geometry: the alignment of celestial bodies. We will therefore search not in time or phase, but rather in orbital parameters space: for a given test planetary (and binary) orbit the projected distances between the planet and each of the stars is known, and occurrence of a transit is exactly true or false at each point in time (ignoring planetary ingress/egress). One can then, similarly to BLS, fit a discrete-valued function to the data - where all the in-transit and out-of-transit points are already separated, using the CB-BLS statistic above (see Fig. \ref{LC_vs_d}). The output is a multi-dimensional "periodogram" - where the peak value corresponds to the best fit not only in orbital period, but also all other tested orbital elements.

More intuitively, it allows to "fold" the LC not in orbital phase, but in projected separation - where, ideally, all points closer to component 1 than R1 will be in-transit (of component 1), and similarly for component 2 (see Fig. 2).

\section{Tests on Simulated Data}

\begin{figure}[b]
\begin{center}
 \includegraphics[width=2.5in]{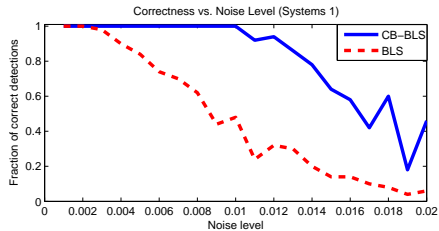} 
\caption{CB-BLS (top) and BLS (bottom) periodograms for the same realization (system with 1\% added white noise. Note the CB-BLS periodogram has no significant aliasing since the model fitted is not periodic.}
\label{Correctness}
\end{center}
\end{figure}

We simulated several systems - starting from 3D 3-body interaction, through LC generation, and to CB-BLS analysis. We varied different system parameters in order to explore the properties of CB-BLS, and each system was realized 50 times with random white noise. The simulated transits were 0.4\% and 0.6\% deep for the transits of the primary and secondary components, respectively. For e.g., Fig. \ref{OnePeriodogram} shows of one CB-BLS periodogram, and Fig. \ref{Correctness} shows the correct identification rate as a function of the amplitude of the (white) noise. Similar results were obtain when we varied the planetary period or the planetary orbital inclination [Ofir 2008].

\section{Conclusions}

CB-BLS is superior to BLS in finding CB planets simply because BLS is not designed to do so.

\underline{A new tool:} The CB-BLS algorithm allows to find transiting CB planets in the residuals of EBs that have noise level of 1\% or more. CB-BLS allows to easily harness existing datasets to the detection of transiting CB planets. False positives fraction is expected to be lower (rare astrophysical false positives). Follow up will be interesting: Orbital evolution on short time scales (100s of days), resonant and/or chaotic systems, four distinct RM effects (1-2, 2-1, p-1, p-2).

\underline{Importance of CB planets:} 1) The frequency of CB planets depends on assumed planetary formation mechanism, and planets in tight systems provide very restrictive constraints on formation models [Muterspaugh \etal (2007)]. 2) Understanding very high precision light curves (e.g., space, TLC project) is limited by stellar
parameters.  EBs allow to significantly increase stellar parameters accuracy. Resultant fits to planetary structure models will have smaller error bars.

\underline{Current status of CB-BLS [beyond Ofir 2008]}
CB-BLS evolved to:
\begin{itemize}
\item Simultaneously fits multi-band data.
\item Naturally includes Roche geometry (e.g. overcontact binaries).
\item Uses the directional correction [Tingley 2003] to further reduce false positives.
\end{itemize}

\end{document}